\documentclass[12pt]{article}
\usepackage{amssymb}
\usepackage[]{hyperref}
\usepackage{graphicx}
\begin{document}
\newcommand{\beq}{\begin{equation}}
\newcommand{\eeq}{\end{equation}}
\newcommand{\beqa}{\begin{eqnarray}}
\newcommand{\eeqa}{\end{eqnarray}}
\newcommand{\beqar}{\begin{eqnarray*}}
\newcommand{\eeqar}{\end{eqnarray*}}
\newcommand{\al}{\alpha}
\newcommand{\be}{\beta}
\newcommand{\del}{\delta}
\newcommand{\D}{\Delta}
\newcommand{\eps}{\epsilon}
\newcommand{\ga}{\gamma}
\newcommand{\Ga}{\Gamma}
\newcommand{\ka}{\kappa}
\newcommand{\nn}{\nonumber}
\newcommand{\inn}{\!\cdot\!}
\newcommand{\h}{\eta}
\newcommand{\ii}{\iota}
\newcommand{\kk}{\varphi}
\newcommand\F{{}_3F_2}
\newcommand{\la}{\lambda}
\newcommand{\La}{\Lambda}
\newcommand{\na}{\prt}
\newcommand{\Om}{\Omega}
\newcommand{\om}{\omega}
\newcommand{\p}{\Phi}
\newcommand{\sig}{\sigma}
\renewcommand{\t}{\theta}
\newcommand{\z}{\zeta}
\newcommand{\ssc}{\scriptscriptstyle}
\newcommand{\eg}{{\it e.g.,}\ }
\newcommand{\ie}{{\it i.e.,}\ }
\newcommand{\labell}[1]{\label{#1}} %{\label{#1}} %
\newcommand{\reef}[1]{(\ref{#1})}
\newcommand\prt{\partial}
\newcommand\veps{\varepsilon}
\newcommand{\pol}{\varepsilon}
\newcommand\vp{\varphi}
\newcommand\ls{\ell_s}
\newcommand\cF{{\cal F}}
\newcommand\cA{{\cal A}}
\newcommand\cS{{\cal S}}
\newcommand\cT{{\cal T}}
\newcommand\cV{{\cal V}}
\newcommand\cL{{\cal L}}
\newcommand\cM{{\cal M}}
\newcommand\cN{{\cal N}}
\newcommand\cG{{\cal G}}
\newcommand\cK{{\cal K}}
\newcommand\cH{{\cal H}}
\newcommand\cI{{\cal I}}
\newcommand\cJ{{\cal J}}
\newcommand\cl{{\iota}}
\newcommand\cP{{\cal P}}
\newcommand\cQ{{\cal Q}}
\newcommand\cg{{\tilde {{\cal G}}}}
\newcommand\cR{{\cal R}}
\newcommand\cB{{\cal B}}
\newcommand\cO{{\cal O}}
\newcommand\tcO{{\tilde {{\cal O}}}}
\newcommand\bz{\bar{z}}
\newcommand\bb{\bar{b}}
\newcommand\ba{\bar{a}}
\newcommand\bg{\bar{g}}
\newcommand\bc{\bar{c}}
\newcommand\bw{\bar{w}}
\newcommand\bX{\bar{X}}
\newcommand\bK{\bar{K}}
\newcommand\bA{\bar{A}}
\newcommand\bH{\bar{H}}
\newcommand\bF{\bar{F}}
\newcommand\bxi{\bar{\xi}}
\newcommand\bphi{\bar{\phi}}
\newcommand\bpsi{\bar{\psi}}
\newcommand\bprt{\bar{\prt}}
\newcommand\bet{\bar{\eta}}
\newcommand\btau{\bar{\tau}}
\newcommand\hF{\hat{F}}
\newcommand\hA{\hat{A}}
\newcommand\hT{\hat{T}}
\newcommand\htau{\hat{\tau}}
\newcommand\hD{\hat{D}}
\newcommand\hf{\hat{f}}
\newcommand\hK{\hat{K}}
\newcommand\hg{\hat{g}}
\newcommand\hp{\hat{\Phi}}
\newcommand\hi{\hat{i}}
\newcommand\ha{\hat{a}}
\newcommand\hb{\hat{b}}
\newcommand\hQ{\hat{Q}}
\newcommand\hP{\hat{\Phi}}
\newcommand\hS{\hat{S}}
\newcommand\hX{\hat{X}}
\newcommand\tL{\tilde{\cal L}}
\newcommand\hL{\hat{\cal L}}
\newcommand\tG{{\tilde G}}
\newcommand\tg{{\tilde g}}
\newcommand\tphi{{\widetilde \Phi}}
\newcommand\tPhi{{\widetilde \Phi}}
\newcommand\te{{\tilde e}}
\newcommand\tk{{\tilde k}}
\newcommand\tf{{\tilde f}}
\newcommand\tH{{\tilde H}}
\newcommand\ta{{\tilde a}}
\newcommand\tb{{\tilde b}}
\newcommand\tc{{\tilde c}}
\newcommand\td{{\tilde d}}
\newcommand\tm{{\tilde m}}
\newcommand\tmu{{\tilde \mu}}
\newcommand\tnu{{\tilde \nu}}
\newcommand\talpha{{\tilde \alpha}}
\newcommand\tbeta{{\tilde \beta}}
\newcommand\trho{{\tilde \rho}}
 \newcommand\tR{{\tilde R}}
\newcommand\teta{{\tilde \eta}}
\newcommand\tF{{\widetilde F}}
\newcommand\tK{{\tilde K}}
\newcommand\tE{{\widetilde E}}
\newcommand\tpsi{{\tilde \psi}}
\newcommand\tX{{\widetilde X}}
\newcommand\tD{{\widetilde D}}
\newcommand\tO{{\widetilde O}}
\newcommand\tS{{\tilde S}}
\newcommand\tB{{\tilde B}}
\newcommand\tA{{\widetilde A}}
\newcommand\tT{{\widetilde T}}
\newcommand\tC{{\widetilde C}}
\newcommand\tV{{\widetilde V}}
\newcommand\thF{{\widetilde {\hat {F}}}}
\newcommand\Tr{{\rm Tr}}
\newcommand\tr{{\rm tr}}
\newcommand\STr{{\rm STr}}
\newcommand\hR{\hat{R}}
\newcommand\M[2]{M^{#1}{}_{#2}}
\newcommand\MZ{\mathbb{Z}}
\newcommand\MR{\mathbb{R}}
\newcommand\bS{\textbf{ S}}
\newcommand\bI{\textbf{ I}}
\newcommand\bJ{\textbf{ J}}

%\begin{document}
\begin{titlepage}
\begin{center}

\vskip 0.5 cm
{\LARGE \bf 
Tachyon couplings from  S-matrix elements\\  \vskip 0.25 cm  in bosonic string theory
} \\
\vskip 1.25 cm
 Mohammad R. Garousi \footnote{garousi@um.ac.ir}

\vskip 1 cm
{{\it Department of Physics, Faculty of Science, Ferdowsi University of Mashhad\\}{\it P.O. Box 1436, Mashhad, Iran}\\}
\vskip .1 cm
 \end{center}

\begin{abstract}
We introduce a systematic framework for constructing spacetime and D-brane effective actions in bosonic string theory, encompassing both tachyon and massless modes. The actions are required to be gauge invariant and compatible with the expansion of sphere- and disk-level S-matrix elements. Our central proposal stipulates that, for each closed- or open-string channel, S-matrix poles involving an odd number of tachyons must be reproduced via an expansion in the effective action, whereas those with an even number of tachyons must be matched exactly. This criterion imposes strong selection rules: the bulk action contains only even powers of closed-string tachyon fields, and the D-brane action only even powers of open-string tachyon fields. In contrast, couplings of closed-string tachyons to D-branes are less constrained and allow both even and odd field multiplicities. As a concrete application, we analyze the disk-level amplitude involving two closed-string tachyons and demonstrate that the resulting linear and quadratic tachyon–D-brane interactions coincide precisely with those of type 0 theory. This is consistent with the duality between the orientifold of type 0 theory and the compactification of bosonic string theory on \(T^{16}\).

\end{abstract}

%Keywords: T-duality, D-brane effective action
\end{titlepage}

\section{The basic idea}

Recent studies suggest that the M-theoretic interpretation of type 0A theory and the F-theoretic interpretation of type 0B theory require the type 0 tachyon to admit a geometric interpretation, relating it to the radii of two circles in the wedge space \(S^1 \vee S^1\) \cite{Baykara:2026gem,Dasgupta:2026maq,Basile:2026trt,Kamal:2026msr}. Moreover, the compactification of bosonic string theory on \(T^{16}\) is conjectured to be dual to the orientifold of type 0 theory \cite{Bergman:1997rf,Dudas:2001wd,Baykara:2026vdc}. These observations motivate a detailed investigation of the effective action of bosonic string theory, with particular emphasis on the tachyon sector.

   One standard method for deriving the spacetime effective action and the D-brane world-volume action in the massless sector is the S-matrix approach \cite{Gross:1986iv,Gross:1986mw}. In this method, the string-theoretic S-matrix elements of massless vertex operators are computed and then expanded at low energy, corresponding to the limit $\alpha'\rightarrow 0$. This expansion must then be reproduced by an effective action that is a series in higher-covariant-derivative couplings of the massless fields. In superstring theories, which are tachyon-free, the S-matrix elements receive contributions from various channels in which both massless and an infinite tower of massive string excitations propagate. The massless poles are retained, while the massive poles are expanded to generate higher-momentum contact terms. The massless poles are reproduced by the leading-order terms of the covariant effective action, whereas the effects of the higher-momentum contact terms manifest as higher-covariant-derivative corrections to the leading-order action. The corresponding effective action can also be obtained via T-duality \cite{Garousi:2017fbe,Garousi:2019wgz,Garousi:2020gio}. In practice, it is considerably easier to derive the effective action by imposing T-duality and to use the S-matrix method only to fix a few remaining parameters that T-duality cannot determine \cite{Ameri:2025bei}. In the case of superstring theory, supersymmetry may also be employed to determine the effective action \cite{Ozkan:2024euj}.

In bosonic string theory, which involves tachyons, the various channels of the S-matrix elements of massless external states receive contributions from tachyon, massless, and an infinite tower of massive states. In this case as well, to obtain a covariant effective action for the massless fields, one must retain the massless poles while expanding the tachyon and massive poles. Indeed, in the four-graviton sphere-level S-matrix element \cite{Schwarz:1982jn} and in the two-graviton disk-level S-matrix element in bosonic string theory \cite{Corley:2001hg}, the massless poles are kept while the tachyon and all massive fields are expanded. The leading-order terms of these expansions are reproduced by covariant effective actions \cite{Metsaev:1986yb,Gholian:2023kjj,Corley:2001hg}. These effective actions can also be derived via T-duality \cite{Garousi:2019wgz,Garousi:2019mca,Ameri:2025bei}. In fact, T-duality fixes the spacetime effective action of bosonic string theory at orders \(\alpha'^0, \alpha'^1, \alpha'^2\) up to one overall factor \cite{Garousi:2019wgz,Garousi:2019mca,Ameri:2025bei}, and the D-brane effective action at orders \(\alpha'^0, \alpha'^1\) up to one overall factor \cite{Garousi:2013gea,Hosseini:2022vrr}. This implies that no alternative expansion exists for the S-matrix elements of massless fields. Since we are interested in effective actions that involve both massless and tachyon fields, one might propose that the effective action should include couplings of two gravitons and one tachyon. Indeed, the sphere-level S-matrix element of two gravitons and one tachyon is non-zero \cite{Schwarz:1982jn}. However, if the effective action includes such a coupling, then the tachyon poles of the S-matrix elements must be reproduced by this coupling. This means that the higher-momentum contact terms—which must be reproduced by higher-derivative terms—are modified relative to the case where the tachyon poles are expanded. Given that there is a unique covariant action consistent with T-duality \cite{Metsaev:1986yb,Gholian:2023kjj,Corley:2001hg}, the resulting higher-momentum terms of the S-matrix element cannot be reproduced by a covariant coupling and are inconsistent with T-duality. Hence, even though the sphere-level S-matrix element of two gravitons and one tachyon is non-zero, the effective action should not include this coupling.

 The same conclusion applies to the disk-level S-matrix element involving massless non-Abelian gauge bosons. These amplitudes contain poles associated with the exchange of open-string tachyons, massless states, and an infinite tower of massive excitations \cite{Schwarz:1982jn,Kawai:1985xq}. Nevertheless, in order to extract gauge-invariant couplings, one must expand the open-string tachyon poles \cite{Neveu:1971mu} (see, also the Appendix of \cite{Garousi:2015qgr}). As a result, the effective field theory should not include a three-point coupling of two gauge bosons and a single open-string tachyon, despite the fact that the corresponding disk-level amplitude with two gauge boson vertex operators and one open-string tachyon vertex operator is nonzero \cite{Schwarz:1982jn}. Furthermore, the gauge-invariant couplings obtained from the expanded tachyon pole are compatible with T-duality \cite{Garousi:2015qgr,Baradaran-Hosseini:2024xap}.

We propose that, even though the sphere-level (or disk-level) S-matrix elements involving an odd number of closed-string (or open-string) tachyons and an arbitrary number of massless vertex operators are nonzero \cite{Schwarz:1982jn}, the corresponding effective action should not include them. Specifically, the spacetime effective action should include only an even number of closed-string tachyons, and the D-brane effective action should include only an even number of open-string tachyons. The rationale is that the S-matrix elements of tachyon and massless states in bosonic string theory contain both tachyon and massless poles \cite{Schwarz:1982jn}; the appropriate expansion keeps one pole and expands the other. Therefore, the effective action, which involves an even number of tachyons, should reproduce only one of these poles. The pole of the S-matrix that is expanded should instead be reproduced by the higher-covariant-derivative terms of the same field theory, which also involve an even number of tachyons. This is in contrast to the S-matrix elements of NS-NS vertex operators in type 0 string theory \cite{Klebanov:1998yya,Garousi:2003db}, where each channel contains either a massless pole or a tachyon pole. Importantly, it is these poles themselves—not their expansion—that must be reproduced by the corresponding field theory. In other words, in the NS-NS sector of type 0 string theory, the S-matrix elements do not require any expansion of massless or tachyon poles. Consequently, all non-vanishing S-matrix elements in type 0 theory should be accounted for directly by the effective action. (See also the Discussion section for a comment on the couplings of two R-R fields and one tachyon.)

    To clarify the above proposal, let us consider the scattering amplitude of $i$ external closed-string tachyons and $j$ external gravitons from a D-brane in bosonic string theory. The amplitude contains, among other contributions, a closed-string channel in which a single closed string propagates between the D-brane and the vertex of the closed-string fields. One may call this the \((p_1+\cdots+p_{i+j})^2\)-channel. If \(i\) is odd, the tachyon pole—which is produced by the D-brane and a bulk coupling of \(i+1\) tachyons (an even number) and \(j\) gravitons—should be reproduced by the field theory, while the massless pole, which is produced by the D-brane and a bulk coupling of \(i\) tachyons (an odd number) and \(j+1\) gravitons, should be expanded. This expansion is taken around \((p_1+\cdots+p_{i+j})^2 \to -m^2\), where \(m\) is the tachyon mass. Hence, the field theory should not include odd-number tachyon couplings. Conversely, if \(i\) is even, the massless pole—which is produced by the D-brane and a bulk coupling of \(i\) tachyons (an even number) and \(j+1\) gravitons—should be reproduced by the field theory, while the tachyon pole, which involves \(i+1\) tachyons (an odd number) and \(j\) gravitons, should be expanded around \((p_1+\cdots+p_{i+j})^2 \to 0\). Thus, the field theory should involve only even numbers of closed-string tachyons.
Notably, the contact terms resulting from the expansion in the first case produce couplings of an odd number of closed-string tachyons to the D-brane, whereas in the second case they produce couplings of an even number.

A similar observation holds for disk-level S-matrix elements involving an odd number of open-string tachyons and non-Abelian gauge bosons. For example, the S-matrix element of one open-string tachyon and three gauge bosons is nonzero and contains both massless and tachyon poles. Since there are no couplings of two gauge bosons and one open-string tachyon in the D-brane effective action, the field theory reproduces neither the massless pole nor the tachyon poles. Consequently, these poles in the string-theory S-matrix element must be expanded. The resulting higher-momentum contact terms are nonzero; however, they should not be included in the effective action, as it already contains higher-momentum four-gauge-boson couplings.
One can easily verify that in the string-theory S-matrix element with two tachyons and three gauge bosons, both massless and tachyon poles are present in each channel. The massless pole should be reproduced by field-theory couplings involving two tachyons and one gauge field, as well as four-gauge-field couplings, while the tachyon pole—which would be reproduced by couplings of three tachyons and couplings of one tachyon and three gauge fields—must be expanded and reproduced by higher-derivative couplings of two tachyons and three gauge fields.

As another example, consider the S-matrix element of three tachyons and one gauge boson. This amplitude again has both tachyon and massless poles; however, the field theory produces neither of them. Hence, both poles must be expanded in the string-theory S-matrix element. The resulting higher-momentum contact terms of one gauge field and three tachyons should also not be included in the effective action, because the effective action is required to include higher-derivative two-tachyon and two-gauge-field couplings. The field theory should include only one of these two couplings, since in the S-matrix elements with one gauge boson and four tachyon vertex operators, the poles again include both massless and tachyon contributions. One of them must be expanded, while the other must be reproduced by field-theory couplings. The tachyon pole corresponding to the vertex of three tachyons and one gauge boson must be expanded to produce higher-derivative couplings of one gauge field and four tachyons, whereas the massless pole corresponding to the vertex of two tachyons and two gauge fields must be reproduced by the field-theory couplings of two tachyons and one gauge field, as well as the couplings of four tachyons. Hence, the field theory should not contain couplings between three tachyons and one gauge boson at any order of derivatives. The same reasoning applies to all higher-point functions. Therefore, we speculate that the covariant spacetime effective action should include only even numbers of closed-string tachyons, whereas D-brane couplings should contain only even numbers of open-string tachyons—with no restriction on the number of closed-string tachyons that couple to the D-brane.

A crucial point is that if one could determine all field-theory couplings that involve only even numbers of tachyons, the S-matrix elements derived from that field theory would exactly reproduce the string-theory S-matrix expansion for even numbers of tachyon vertex operators. Consequently, the field theory would automatically incorporate the effects of all massive states and odd-number tachyons, as these contributions are already contained in the string-theory S-matrix elements of the even-number tachyon vertices.

A similar expansion for the S‑matrix element of an even number of tachyons has been proposed in \cite{Garousi:2002wq,Garousi:2003ur,Bitaghsir-Fadafan:2006iya,Garousi:2003db}. The expanded S‑matrix elements should be such that they reproduce the field‑theory poles arising from the standard covariant kinetic terms of the tachyon and the massless fields. This expansion, however, does not address the S‑matrix element of an odd number of tachyons. The expansion proposed in the present paper is consistent with the method of \cite{Garousi:2002wq,Garousi:2003ur,Bitaghsir-Fadafan:2006iya,Garousi:2003db} and indicates that there is no coupling of odd numbers of tachyons in the covariant effective action.

 In the next section, we apply the above expansion to the disk-level S-matrix element of two closed-string tachyons in bosonic string theory, with the aim of deriving the linear and quadratic couplings of the closed-string tachyon to D-branes. We show that these couplings are exactly identical to the corresponding tachyon couplings in type 0 theory \cite{Klebanov:1998yya,Garousi:1999fu}. This confirms the conjectured duality between bosonic string theory on \(T^{16}\) and the orientifold of type 0 theory \cite{Bergman:1997rf,Dudas:2001wd,Baykara:2026vdc}. In Subsection 2.2, we examine the amplitude involving one tachyon and one dilaton. The expansion of this amplitude reproduces the field-theory couplings exactly, thereby confirming the prescription for expanding the S-matrix elements. A brief discussion of our results and their implications is provided in Section 3.

\section{Application to  specific S-matrix elements}

In this section, we apply the prescription introduced in this paper to analyze the scattering amplitudes involving two tachyons, as well as one tachyon and one dilaton, on a D\(_p\)-brane within the bosonic string theory. The analysis of the first amplitude yields both linear and quadratic corrections to the D-brane tension, which we find to be identical to those obtained in the type 0 theory \cite{Klebanov:1998yya,Garousi:1999fu}. The examination of the second amplitude independently corroborates the couplings derived from the two-tachyon amplitude. Since both amplitudes reproduce the expected couplings, they provide strong support for the proposed prescription for expanding S-matrix elements involving tachyons.

\subsection{Two-tachyon amplitude}

    The disk-level S-matrix element for two tachyons was previously computed in type 0 theory in \cite{Garousi:1999fu}. That amplitude is free of tachyon poles, which renders its low-energy expansion and the determination of the associated tachyon couplings to the D$_p$-brane relatively straightforward. The corresponding computation in the bosonic theory, by contrast, gives rise to both massless and tachyon poles in all channels. In this section, we adopt the prescription developed in this paper to study this scattering amplitude in the bosonic string theory.

 The relevant world-sheet quantity is given by the following disk-level S-matrix element of two closed-string vertex operators:
 \beqa
 A &\sim& \langle Q_1 Q_2 \rangle \,,\labell{A}
 \eeqa 
 where each vertex operator is
 \beqa
 Q_i &=& \int d^2z_i \, :e^{ip_i\cdot X(z_i)}:\,:e^{ip_i\cdot D\cdot X(\bar{z}_i)}:\,. \labell{Qtachyon}
 \eeqa
 The tachyon momenta satisfy the on-shell relation \(p_i^2 = 4/\alpha'\). Adopting the convention \(\alpha' = 2\), this reduces to \(p_i^2 = 2\). We employ the doubling trick  \cite{Garousi:1996ad} to express the antiholomorphic part of the world-sheet field, \(\bar{X}^\mu\), in terms of the holomorphic field \(X^\mu\), which is implemented by the matrix \(D_{\mu}{}^{\nu} = \mathrm{diag}(\underbrace{1,1,\dots, 1}_{p+1}, -1,-1,\dots,-1)\). The world-sheet propagator is given by
 \beqa
 \langle X^{\mu}(z) X^{\nu}(w) \rangle &=& -\eta^{\mu\nu} \log(z-w)\,.
 \eeqa
 Using this propagator, one can evaluate the correlator in \reef{A} and show that the resulting integrand is invariant under \(SL(2,\mathbb{R})\). Fixing this symmetry by setting \(z_1 = i\) and \(z_2 = iy\), which transforms the measure as \(\int d^2z_1 d^2z_2 \rightarrow \int_0^1 (1-y^2)\,dy\), and then changing variables to \(y = \frac{1-\sqrt{x}}{1+\sqrt{x}}\), one finds
 \beqa
 A &=& \alpha \int_0^1 dx \, x^{\,p_1\cdot p_2} (1-x)^{\,p_2\cdot D\cdot p_2} \, \delta^{(p+1)}(p_1 + D\cdot p_1 + p_2 + D\cdot p_2)\,,
 \eeqa
 where we have absorbed an overall normalization factor into the constant \(\alpha\). We will fix this constant by comparing the amplitude with its counterpart in field theory. Defining the closed-string Mandelstam variable \(t = -(p_1 + p_2)^2\) and the open-string Mandelstam variable \(s = -p_{1a} p_1^a\), the above amplitude can be rewritten as
 \beqa
 A &=& \alpha \, B(-1 - t/2, \, -1 - 2s)\,,
 \eeqa
 where we have suppressed the delta function imposing momentum conservation along the D\(_p\)-brane.

 The amplitude exhibits both closed-string and open-string poles at
 \beqa
 -1 - t/2 = 0,- 1, -2, \dots; \qquad -1 - 2s = 0, -1, -2, \dots\,.
 \eeqa
 As expected, both channels contain poles corresponding to the propagation of tachyons, massless states, and an infinite tower of massive states. Since the external states are two closed-string tachyons, the prescription introduced in this paper dictates that the tachyon and massive poles must be expanded, while the massless poles are to be retained. Consequently, the expansion is carried out in the limit \(s, t \rightarrow 0\).

 Using the identity \(x\Gamma(x) = \Gamma(1+x)\), one can rewrite the amplitude as
 \beqa
 A = -\alpha \, \frac{(-2 - t/2 - 2s)(-t/2 - 2s)}{s t} \left[ \frac{(1+ t/2+ 2s)\Gamma(1 - t/2)\Gamma(1 - 2s)}{(-1 - t/2)(-1 - 2s)\Gamma(1 - t/2 - 2s)} \right]\,.\labell{A3}
 \eeqa
 The bracket contains the tachyon and massive poles, which reduce to unity at leading order in the limit \(s, t \rightarrow 0\). Using the on-shell relations, one finds \(-2 - t/2 - 2s = -s + p_{1i} p_2^i\), so the amplitude can be expressed as
 \beqa
 A = -2\alpha \left( \frac{2 + 2s}{t} - \frac{p_{1i} p_2^i}{4s} + \frac{3}{4} \right) \Big[ 1 + \cdots \Big]\,. \labell{Aexpand}
 \eeqa
 The ellipsis denotes terms arising from the expansion of the tachyon and massive poles. Based on the on-shell relations, we expect these terms to correspond to higher-derivative and T-duality invariant couplings of two tachyons to the D-brane. Since these contributions are not the focus of the present work, we disregard them.

 We now reproduce this amplitude using the appropriate covariant and T-duality invariant bulk and D-brane effective actions. The bulk action in the string frame, at leading order and including the two-tachyon terms, is standard:
 \beqa
 S_{\rm bulk} &=& \frac{1}{\kappa^2}\int d^{26}x \, \sqrt{-g} \, e^{2\Phi} \left[ 2\left( R + 4\partial_\mu \Phi \partial^\mu \Phi - \frac{1}{12}H^2 \right) - \frac{1}{2}\partial_\mu T \partial^\mu T + T^2 \right]\,. \labell{Sbulk}
 \eeqa
 The D-brane action is given by\footnote{Our index convention is that \(\mu, \nu, \cdots\) represent the spacetime indices, \(a, b, \cdots\) represent the world-volume indices, and \(i, j, \cdots\) represent the transverse space indices.}
 \beqa
 S_{\rm brane} &=& -T_p \int d^{p+1}\sigma \, f(T) e^{-\Phi} \sqrt{-\det\left( [g+B]_{ab} + 4\pi F_{ab} \right)}\,, \labell{Sbrane}
 \eeqa
 where \([g]_{ab} = \partial_a X^\mu \partial_b X^\nu g_{\mu\nu}\) is the pullback of the bulk metric to the D\(_p\)-brane, and similarly for \([B]_{ab}\). The function \(f(T)\) is expanded as
 \beqa
 f(T) = 1 + a_1 T + a_2 T^2 + \cdots\,,\labell{fT}
 \eeqa
 with \(a_1\) and \(a_2\) as yet undetermined constants. We will fix these constants by comparing the S-matrix element for two tachyons derived from the above actions with the expansion in \reef{Aexpand}.

 The closed-string fields appearing in the D-brane action are functionals of the string coordinate \(X^\mu\) and must be Taylor-expanded \cite{Garousi:1998fg}. In the static gauge, where \(X^a = \sigma^a\), the expansion of a closed-string field takes the form
 \beqa
 T(X^\mu) = T(\sigma^a) + X^i \partial_i T(\sigma^a) + \cdots\,,
 \eeqa
 where the ellipsis denotes higher-order terms in the Taylor expansion that are not relevant to our calculation. The closed-string tachyon is invariant under T-duality; consequently, the actions in \reef{Sbulk} and \reef{Sbrane} are T-duality invariant \cite{Garousi:2019wgz,Garousi:2017fbe}. In order to reproduce the S-matrix element in \reef{Aexpand} from field theory, we do not need to include the \(B\)-field or the open-string gauge field \(F\). Henceforth, we therefore ignore these fields in the actions.

 The S-matrix elements in string theory should be reproduced by the effective actions in the Einstein frame. We therefore transform to the Einstein frame using the relation \(g_{\mu\nu} = e^{\gamma\Phi} G_{\mu\nu}\), where \(\gamma = 4/(D-2)\). In this frame, the actions take the form
 \beqa
 S_{\rm bulk} &=& \frac{1}{\kappa^2}\int d^{26}x \, \sqrt{-G} \left[ 2\left( R - \gamma \partial_\mu \Phi \partial^\mu \Phi \right) - \frac{1}{2}\partial_\mu T \partial^\mu T + e^{\gamma\Phi} T^2 \right] \,,\nn\\
 S_{\rm brane} &=& -T_p \int d^{p+1}\sigma \, f(T) e^{-\Phi + \frac{p+1}{2}\gamma\Phi} \sqrt{-\det[G]_{ab}}\,. \labell{Sbulkbrane}
 \eeqa
 To obtain standard kinetic terms for the fluctuations, we normalize the fields as
 \beqa
 G_{\mu\nu} = \eta_{\mu\nu} + \kappa h_{\mu\nu}, \qquad \Phi = \frac{\kappa}{2\sqrt{\gamma}} \phi, \qquad T = \kappa \tau, \qquad X^i = \frac{1}{\sqrt{T_p}} \lambda^i\,.
 \eeqa
 The \(t\)-channel contribution in field theory is given by
 \beqa
 A'_t(\tau_1,\tau_2) &=& i\tilde{S}_\phi \tilde{G}_\phi \tilde{V}_{\phi \tau_1\tau_2} + i(\tilde{S}_h)^{ab} (\tilde{G}_h)_{ab,\mu\nu} (\tilde{V}_{h \tau_1\tau_2})^{\mu\nu}\,, \labell{At}
 \eeqa
 where the sources are read off from the D-brane action, while the propagators and vertices are extracted from the bulk action. They are given by
 \beqa
 \tilde{S}_{\phi} &=& -T_p \left[ \frac{p+1}{2}\gamma - 1 \right] \frac{\kappa}{2\sqrt{\gamma}}, \qquad \tilde{S}_h^{ab} = -\frac{1}{2}T_p \kappa \eta^{ab}, \nn\\
 \tilde{G}_\phi &=& -\frac{i}{k^2}, \qquad (\tilde{G}_h)_{ab,\mu\nu} = -\frac{i}{2k^2} \left( \eta_{a\mu}\eta_{b\nu} + \eta_{a\nu}\eta_{b\mu} - \frac{2}{D-2}\eta_{ab}\eta_{\mu\nu} \right), \nn\\
 \tilde{V}_{\phi \tau_1\tau_2} &=& i\kappa\sqrt{\gamma}, \qquad (\tilde{V}_{h \tau_1\tau_2})^{\mu\nu} = -\frac{i\kappa}{2} \left[ p_1^\mu p_2^\nu + p_2^\mu p_1^\nu - \eta^{\mu\nu} (p_1\cdot p_2 + 2) \right]\,.
 \eeqa
 Substituting these expressions into \reef{At}, one finds
 \beqa
 A'_t(\tau_1,\tau_2) = -\frac{iT_p\kappa^2}{4t} (2 + 2s)\,.
 \eeqa
 The \(s\)-channel pole in field theory is given by
 \beqa
 A'_s(\tau_1,\tau_2) &=& (\tilde{V}_{\tau_1\lambda})^i (\tilde{G}_\lambda)_{ij} (\tilde{V}_{\lambda \tau_2})^j\,.
 \eeqa
 The transverse scalar propagator and the vertex for one tachyon and one scalar can be read off from the brane action in \reef{Sbulkbrane}. They are
 \beqa
 (\tilde{G}_\lambda)_{ij} = -\frac{i\eta_{ij}}{k_a k^a}, \qquad (\tilde{V}_{\tau_1\lambda})^i = a_1 \kappa \sqrt{T_p} \, p_1^i\,.\labell{tV}
 \eeqa
 Consequently, the \(s\)-channel pole becomes
 \beqa
 A'_s(\tau_1,\tau_2) &=& iT_p\kappa^2 a_1^2 \, \frac{p_{1i} p_2^i}{s}\,.
 \eeqa
 The contact term for two tachyons can likewise be read off from the brane action, yielding
 \beqa
 A'_c(\tau_1,\tau_2) &=& -2iT_p\kappa^2 a_2\,.
 \eeqa
 The constants \(a_1\) and \(a_2\) are those introduced in the function \(f(T)\) in \reef{fT}.

 Adding the pole and contact contributions in field theory, one obtains the following amplitude:
 \beqa
 A'_t + A'_s + A'_c = -\frac{iT_p\kappa^2}{4} \left[ \frac{2+2s}{t} - 16 a_1^2 \frac{p_{1i}p_2^i}{4s} + 8a_2 \right]\,.
 \eeqa
 Comparing this with the string amplitude in \reef{Aexpand}, we find the normalization of the string amplitude to be \(\alpha = iT_p\kappa^2/8\), and the tachyon function to be
 \beqa
 f(T) = 1 + \frac{T}{4} + \frac{3}{32}T^2 + \cdots\,.\labell{fT2}
 \eeqa
 This is precisely the same function that appears in the D-brane action of type 0 theory \cite{Garousi:1999fu}. Thus, the closed-string tachyon corrections to the tension of the D\(_p\)-brane in bosonic string theory and in type 0 theory are identical. This result is expected based on the conjectured duality between type 0 and bosonic string theories \cite{Bergman:1997rf,Dudas:2001wd,Baykara:2026vdc}. Moreover, the above result confirms the prescription proposed in this paper for expanding the S-matrix element in bosonic string theory. In the next subsection, we further corroborate both the above tachyon function and the expansion prescription by analyzing the S-matrix element of one tachyon and one dilaton.
 
Before concluding this subsection, we point out that the numerator of the massless \(s\)-channel pole in the string amplitude should not contain higher-momentum terms, since the linear coupling of the tachyon to the D-brane receives no derivative corrections. Consequently, there are no higher-momentum corrections to the field theory vertex \(\tilde{V}_{\phi\tau_1\tau_2}\) in \reef{tV}. Hence, the expansion of the terms inside the bracket in \reef{A3} must be proportional to \(s\). The expansion of the Gamma functions begins at order \(ts\), and the remaining part should be expressed as
\[
\frac{1+t/2+2s}{(-1-t/2)(-1-2s)}=\frac{1}{1+2s}+\frac{2s}{(-1-t/2)(-1-2s)}\,,
\]
which is then expanded in the limit \(s,t\rightarrow 0\), yielding higher-momentum contributions proportional to \(s\). Therefore, the amplitude in \reef{A3} possesses no massless \(s\)-channel pole other than the term \(p_{1i}p_2^i/s\), which arises from the leading-order field theory.

\subsection{Tachyon–dilaton amplitude}

 The scattering amplitude for one tachyon and one dilaton is given by the world-sheet correlator in \reef{A}, where the tachyon vertex operator is given by \reef{Qtachyon} and the dilaton vertex operator takes the form
 \beqa
 Q_2 &=& (\epsilon_2 \cdot D)_{\mu\nu} \int d^2z_2 \, :\partial_{z_2} X^\mu(z_2) e^{ip_2\cdot X(z_2)}:\,:\partial_{\bar{z}_2} X^\nu(\bar{z}_2) e^{ip_2\cdot D\cdot X(\bar{z}_2)}:\,,
 \eeqa
 with \(p_2^2 = 0\). The dilaton polarization tensor is given by
 \beqa
 \epsilon_2^{\mu\nu} &=& \frac{1}{\sqrt{D-2}} \left( \eta^{\mu\nu} - p_2^\mu \ell_2^\nu - p_2^\nu \ell_2^\mu \right)\,,
 \eeqa
 where $D$ is the dimension of spacetime, and the auxiliary vector \(\ell_2^\mu\) satisfies \(p_2 \cdot \ell_2 = 1\).

Performing the correlators, one obtains the following result:
 \beqa
 A &=& \beta \Big[ \left( \Tr(\epsilon_2 \cdot D) - 2p_1 \cdot \epsilon_2 \cdot D \cdot p_1 \right) B(-t/2, -1 - 2s) \nn\\&& - p_1 \cdot \epsilon_2 \cdot p_1 \, B(-1 - t/2, -1 - 2s) - p_1 \cdot D \cdot \epsilon_2 \cdot D \cdot p_1 \, B(1 - t/2, -1 - 2s) \Big]\,,
 \eeqa
 where \(\beta\) is an overall normalization constant. This amplitude is valid for the graviton, for which the polarization tensor is symmetric, and for the Kalb–Ramond field, for which the polarization tensor is antisymmetric. For the dilaton, which is our case of interest, after substituting the corresponding polarization tensor into the above amplitude, the auxiliary field cancels, and the final result is
 \beqa
 A &=& -\frac{\beta}{\sqrt{D-2}} \Big[ (D - 2p - 8 - 4s) B(-t/2, -1 - 2s) \nn\\&& + 2B(-1 - t/2, -1 - 2s) + 2B(1 - t/2, -1 - 2s) \Big]\,.\labell{Atp}
 \eeqa
 As can be seen, the amplitude contains tachyon, massless, and an infinite tower of massive poles in both the open-string \(s\)-channel and the closed-string \(t\)-channel. Since the external states include one closed-string tachyon, our prescription for expanding S-matrix elements requires that the tachyon pole in the \(t\)-channel be kept, while the massless and massive poles are to be expanded. Moreover, since there is no open-string tachyon among the external states, the massless pole in the \(s\)-channel must be kept, whereas the open-string tachyon and massive poles are to be expanded. The expansion is therefore carried out around \(t', s \rightarrow 0\), where \(t' = t + 2\). At leading order, this expansion yields
 \beqa
 \frac{\Gamma(1 - t'/2)\Gamma(1 - 2s)}{\Gamma(1 - t'/2 - 2s)} = 1 + \zeta(2) st'+\cdots, \qquad \frac{1}{-1 - 2s} = -1 +2s+ \cdots\,.
 \eeqa
 To apply the above expansion, we use the identity \(x\Gamma(x) = \Gamma(1+x)\) to rewrite the amplitude \reef{Atp} as
 \beqa
 A &=& -\frac{\beta}{\sqrt{D-2}} \frac{\Gamma(1 - t'/2)\Gamma(1 - 2s)}{\Gamma(1 - t'/2 - 2s)} \Big[ (D - 2p - 4) \frac{-t'/2 - 2s}{(-1 - 2s)(-2s)} \nn\\&& \qquad\qquad\qquad\qquad\qquad + \frac{t'/2 + 2s}{s} - \frac{2(t'/2 + 2s)}{t's} + \frac{2(1 - t'/2)}{(-1 - 2s)(-2s)} \Big]\,,
 \eeqa
 where we have also used the identity
 \beqa
 \frac{-1-t'/2-2s}{(-1-2s)(-t'/2)}&=&-\frac{2}{t'}-\frac{1}{1+2s}\,.
 \eeqa
 Expanding the open-string tachyon pole yields
\beqa
 A &=& -\frac{\beta}{\sqrt{D-2}} \frac{\Gamma(1 - t'/2)\Gamma(1 - 2s)}{\Gamma(1 - t'/2 - 2s)} \Big[- (D - 2p - 4) \frac{(-t'/2 - 2s)}{(-2s)}(1-2s+\cdots) \nn\\&& \qquad\qquad\qquad\qquad\qquad - \frac{4}{t'} +2- \frac{(1-t'/2)}{s} + \frac{(1 - t'/2)}{s}(1-2s+\cdots)  \Big]\,.\labell{A4}
 \eeqa
Observe that, as expected, the massless \(s\)-channel pole contains no higher-momentum corrections. By applying the on-shell relation \(-t'/2-s = p_{1i}p_2^i\), the amplitude simplifies to
\[
A = \frac{\beta}{\sqrt{D-2}} \left[ \frac{4}{t'} - (D - 2p - 4) \frac{p_{1i} p_2^i}{2s} + \frac{1}{2}(D - 2p - 4) + \cdots \right]\,,
\]
where the ellipsis denotes higher-momentum contributions. It is important to note that the contact terms displayed above arise precisely from expanding the open-string tachyon pole. In particular, the constant \(2\) appearing in the second term of the second line in \reef{A4} is canceled by the \(-2\) coming from the last term in the same line, which itself originates from the expansion of the tachyon pole.

We now compute the same S-matrix element using the field theory action in \reef{Sbulkbrane}, with the tachyon function given in \reef{fT2}. In field theory, the \(t\)-channel contribution is
 \beqa
 A'_t(\tau_1,\phi_1) &=& i\tilde{S}_{\tau} \tilde{G}_{\tau} \tilde{V}_{\tau \tau_1 \phi_2}\,,
 \eeqa
 where \(\tilde{S}_{\tau} = -T_p \kappa / 4\), \(\tilde{G}_{\tau} = -i/(k^2 - 2)\), and \(\tilde{V}_{\tau \tau_1 \phi_2} = i\sqrt{\gamma}\,\kappa\). The amplitude therefore becomes
 \beqa
 A'_t(\tau_1,\phi_1)  &=& i \frac{T_p \kappa^2}{2\sqrt{D-2} \, t'}\,.
 \eeqa
 The \(s\)-channel contribution is given by
 \beqa
 A'_s(\tau_1,\phi_1) &=& (\tilde{V}_{\tau_1 \lambda})^i (\tilde{G}_{\lambda})_{ij} (\tilde{V}_{\lambda \phi_2})^j\,,
 \eeqa
 which yields
 \beqa
 A'_s (\tau_1,\phi_1) &=& i \frac{T_p \kappa^2 (-D + 2p + 4) \, p_{1i} p_2^i}{16\sqrt{D-2} \, s}\,,
 \eeqa
 and the contact term is
 \beqa
 A'_c (\tau_1,\phi_1) &=& -i \frac{T_p \kappa^2}{16\sqrt{D-2}} (-D + 2p + 4)\,.
 \eeqa
 Adding the pole contributions and the contact term, the field theory amplitude is found to be
 \beqa
 A' &=& i \frac{T_p \kappa^2}{8\sqrt{D-2}} \left[ \frac{4}{t'} - (D - 2p - 4) \frac{p_{1i} p_2^i}{2s} + \frac{1}{2}(D - 2p - 4) \right]\,.
 \eeqa
 This matches the string amplitude upon fixing the normalization to \(\beta = i T_p \kappa^2 / 8\), which is the same normalization obtained for the two-tachyon amplitude. This result independently confirms both the tachyon function in \reef{fT2} at linear order and our prescription for expanding the S-matrix element.

\section{Conclusion}

 In this paper, we have proposed a method for expanding S-matrix elements in bosonic string theory, where each physical channel contains massless and tachyon poles, as well as an infinite tower of massive poles. The prescription is as follows: massless or tachyon poles whose field-theoretic construction requires couplings involving an odd number of tachyons are to be expanded, whereas those requiring an even number of tachyons are to be retained and reproduced by an effective action that includes massless fields and only even numbers of tachyons. We argue that couplings with an odd number of tachyons are inconsistent with a covariant effective action and, therefore, should not be included in it.
 
 We then apply this prescription to study the expansion of the disk-level S-matrix element of two closed-string tachyons. The resulting field theory, consistent with this expansion, yields linear and quadratic tachyon couplings to the D\(_p\)-brane. These couplings exhibit the same functional dependence as those appearing on D\(_p\)-branes in type 0 string theory \cite{Garousi:1999fu}, where the corresponding S-matrix elements contain either massless or tachyon poles, but not both.
We further employ this prescription to examine the expansion of the disk-level S-matrix element of one tachyon and one dilaton. In this case also, we find that the leading-order terms of the expansion are correctly reproduced by the field theory.
Having the same tachyon couplings to D-branes in type 0 and in the bosonic string theory, is consistent with the conjectured that bosonic string theory on $T^{16}$ and the orientifold of type 0 theory are dual to each other \cite{Bergman:1997rf,Dudas:2001wd,Baykara:2026vdc}. 

 Our analysis of the two-tachyon disk-level S-matrix element allows us to extract the tachyon function \(f(T)\) in \reef{fT2} up to quadratic order. A conjecture put forward in \cite{Garousi:1999fu} suggests that this function is given by \(1/\sqrt{1 - T/2}\). To determine the cubic tachyon couplings within this function, one would need to evaluate the three-tachyon disk amplitude and expand it following the prescription introduced in this work. We leave this detailed computation for future investigation.

 Our detailed analysis of the S-matrix elements considered here shows that the leading-order terms in their expansion are reproduced by the leading-order spacetime and D-brane covariant effective action \reef{Sbulkbrane}, which is also T-duality invariant. We expect similar behavior for other S-matrix elements. The subleading terms should be reproduced by higher-covariant-derivative couplings that are likewise T-duality invariant. Due to on-shell ambiguities in higher-momentum contact terms—in particular, the ambiguity that \(\partial_\mu \partial^\mu T = -2T\)—one must first construct a covariant and T-duality-invariant effective action involving higher derivatives of tachyon and massless fields, and then fix the remaining T-duality-free parameters via S-matrix matching. While T-duality strongly constrains the massless-field action up to a few parameters fixed by S-matrix elements \cite{Garousi:2019wgz,Garousi:2019mca,Ameri:2025bei}, the T-duality invariance of the tachyon suggests that more parameters will remain undetermined when tachyon–massless couplings are included. For example, the tachyon function \(f(T)\)  in \reef{fT2} cannot be fixed by T-duality alone. Nevertheless, S-matrix expansions should suffice to fix all of them. It would be worthwhile to pursue a combined T-duality and S-matrix approach to determine the spacetime tachyon effective action at four- and higher-derivative orders.

A detailed analysis of the sphere-level S-matrix element for four tachyons, as well as the S-matrix element for two tachyons and two massless NS-NS states in type 0 string theory—which contains either massless or tachyon poles—indicates the absence of couplings of the form \(T^4\), \(T^2(\partial T)^2\), \(T^2 R\), \(T^2 H^2\), or \(T^2(\partial \Phi)^2\) \cite{Garousi:2003db}. In other words, the bulk action \reef{Sbulk} in type 0 theory does not contain an overall \(T^2\) factor. This conclusion is consistent with the fact that the type 0 effective action includes a mass term \(T^2\), and under the field redefinition \(T \rightarrow T + \delta T\), it generates a \(T \delta T\) term. Since the NSNS sector effective action contains only even powers of tachyons, the higher-derivative field redefinition \(\delta T\) must involve an odd number of tachyons. Therefore, an \(\alpha'\)-order redefinition of the form \(\delta T = b_1 T R + b_2 T H^2 + \cdots\) can cancel the \(T^2 R\) and \(T^2 H^2\) terms in the effective action. In the bosonic string theory, where a mass term \(T^2\) is also present, the same field redefinition is expected to remove analogous \(T^2 R\) and \(T^2 H^2\) contributions.
 It would therefore be interesting to investigate the corresponding sphere-level S-matrix elements in bosonic string theory, where each channel contains both tachyon and massless poles. Using our prescription for expanding the amplitude, one can explicitly check whether a \(T^2\) factor appears in the bulk action \(\reef{Sbulk}\) in the bosonic string context.

We have seen that although the S-matrix element involving an odd number of tachyons and an arbitrary number of massless fields is nonzero in bosonic string theory, such terms should not be included in the effective action. In type 0 theory, the S-matrix element for one tachyon and two massless R-R states of opposite chirality is also nonzero \cite{Klebanov:1998yya}. In the literature \cite{Klebanov:1998yya}, this coupling is included in the effective action, which subsequently leads—through a detailed analysis of the S-matrix element involving two R-R states and two tachyons—to a coupling of the form \(F^2 T^2\) as well.
However, as in the bosonic case, one may consistently exclude the \(F\bar{F}T\) coupling from the effective action. A justification for this omission is that the S-matrix element of four massless R-R vertex operators with opposite chirality contains a tachyon pole (see the Appendix in \cite{Garousi:2003db}). Since all external states are massless, one expects the amplitude to be expanded in the limit \(\alpha' \to 0\), which implies that the tachyon pole itself must be expanded rather than kept as a singularity. Consequently, the field theory should not include the \(F\bar{F}T\) coupling. It would therefore be interesting to construct a type 0 effective action that consistently excludes such a term from the outset.

%Therefore, the field theory should not include the \(F\bar{F}T\) coupling. We then expect the leading-order effective action to take the following form:
 %\beqa
% S_{\rm type 0} &=& \frac{1}{\kappa^2}\int d^{10}x \, \sqrt{-g} \, e^{2\Phi} \left[ 2\left( R + 4\partial_\mu \Phi \partial^\mu \Phi - \frac{1}{12}H^2 \right) - \frac{1}{2}\partial_\mu T \partial^\mu T + \frac{1}{2}T^2 \right]\nn\\ &&
%  -\frac{1}{\kappa^2}\int d^{10}x \, \sqrt{-g} \left[\frac{1}{2}\sum_{n=1}^4(F_{(n)}\cdot F_{(n)}+\bar{F}_{(n)}\cdot \bar{F}_{(n)})+\frac{1}{2}F_{(5)}F_{(5)}\right]\,,
% \eeqa
% where the R-R terms with \(n=2,4\) correspond to type 0A, whereas the remaining R-R terms correspond to type 0B theory.

%\newpage

\end{document}